\documentclass[epj]{svjour}
\usepackage{graphicx}
\usepackage{bm}
\usepackage{hyperref}
\usepackage{textcomp}
\begin{document}
\title{Efficiency of photodesorption of Rb atoms collected on polymer organic film in vapor-cell}
\author{Sergey N. Atutov\thanks{\email{atutov@fe.infn.it}}
\and Viacheslav P. Chubakov \and Pavel A. Chubakov \and Alexander I. Plekhanov}

\institute{Institute of Automation and Electrometry Sib. RAS, Koptuga 1, 630090 Novosibirsk 90, Russia}
\date{Received: date / Revised version: date}
%
\abstract{
The efficiency of photodesorption of Rb atoms previously collected on polymer organic film has been studied in detail. This study was carried out in a Pyrex glass cell of which the inner surface was covered with (poly)dimethylsiloxane (PDMS) film and illuminated by a powerful flash lamp. The desorption dynamic of the Rb atoms density in the cell caused by the illumination and percentage of desorbed atoms was studied by using of Rb resonance lamp and free running diode laser as sources of probing light. It was determined that 85 percent collected chemical active Rb atoms and stored during 16 seconds in the closed cell, 75 percent in the pumped cell can be desorbed by single flash of the lamp. The number of stored atoms decays with a characteristic time of 60 min in isolated cell and with a time 12.4 minutes in a pumped cell. We believe that this efficient method of collection and fast realization of atoms or molecules could be used for enhancement of sensitivity of existed sensors for the trace detection of various elements (including toxic or radioactive ones) which is important to environmental applications, medicine or in geology. The effect might help to construct an efficient light-driven atomic source for a magneto-optical trap in a case of extremely low vapor density or very weak flux of atoms, such as artificial radioactive alkali atoms.
} 
\maketitle
%
\section{Introduction}
\label{sec:1introduction}
%
Among of the processes of the interaction of light with solids, that of the photodesorption of particles such as atoms or molecules from a solid surface is very important. According to numerous publications devoted to the study of the effect, the photodesorption is a phenomenon in which the particles are released from a different types solid surfaces in a variety of ways by illuminating of the surfaces with sources of visible or UV light (see, for example, Ref. \cite{Bonch1993,Balzer1997} - atomic desorption, Ref. \cite{Bejan2004} - molecular desorption, and references therein). The effect occurs in a system in the state of sorption equilibrium between a gas phase and an adsorbing surface. The photodesorption can be a thermal process due to a direct heating of the surface by an incident light or due to various non-thermal effects. The photodesorption process is characterized by a desorption rate $R$, which, in general, is a function of the adsorption the energy of the particles to the surface, temperature, wavelength of the desorbing light and structure of the solid surface.

It is important to note that in a case of a long time illumination of the surface by light, the photodesorption rate and, thus, photodesorption yield of the particles decreases from maximum to zero because of a decrease of number of adsorbed particles on the surface, which is cleaned by the light. In this case, the photodesorption yield decay curve demonstrates a pure exponential form which is an indication that only the solid surface is involved in desorption process.

The diffusion of atoms or molecules in a bulk within to a surface onto which the particles have been absorbed can also be involved in photodesorption process. As far as we are aware, the first time the important contribution of particle diffusion in bulk to the photodesorption yield from the surface was recognized and was taken into account in an experiment performed on Rb atoms absorbed onto transparent non-stick organic film \cite{Atutov1999}. In this experiment it was shown that when the diffusion of the particles is important, the bulk of the coating together with the coating surface can act as a container for a large number of stored particles. If the adsorption energy of atoms or molecules to the surface is small, these particles can be desorbed by a weak light. Because of the photodesorption, the density of particles in a bulk within to the surface of polymer organic film is decreased to zero. This leads to a transportation of the particles from the inside of the coating towards the surface, where, finally, the desorption from the coating to the vapor phase is accomplished.

Under continuous illumination or pumping of the adsorbed particles from the surface, the desorption yield demonstrates more or less fast decay curve which is followed by a long diffusion tail. This form of the curve is in contrast to the exponential decay curve in the case of the pure surface desorption considered above. This can be one of the indications that the particle diffusion in the bulk is involved in desorption process \cite{Atutov1999}.

The first time this effect consisting of a huge emission of alkaline atoms from siloxane film, under illumination by a laser or an ordinary light source was observed and experimentally studied with Na immersed in polydimethylsiloxane (PDMS) film \cite{Gozzini1993}. Later this effect was observed using a wide range of surfaces and objects, silane-coated glass, in particular, PDMS coating with Rb \cite{Meucci1994}, Cs \cite{Mariotti1996}, Na as well as Na$_{2}$ molecules \cite{Xu1996}, with K \cite{Gozzini2004} and Ca atoms \cite{Mango2008}, octadecyltrichlorosilane (OTS) Rb \cite{Cappello2007}, paraffin with Cs and Rb \cite{Alexandrov2002}, paraffin with Na \cite{Gozzini2008}, superfluid 4He film with Rb \cite{Hatakeyama2000}. Rb and Cs  photodesorption from porous silica have been observed in \cite{Burchianti2004}. Rubidium light-induced desorption from an octadecyldimethylmethoxysilane (ODMS) coating within a photonic band-gap fiber was demonstrated in \cite{Ghosh2006}. Recently results of study of diffusion and photodesorption of Rb from porous alumina were presented in \cite{Villalba2010}.

In practice, the experimental situation can smoothly vary from being one of pure surface desorption (for example, photodesorption from clean sapphire, quartz or metallic surfaces) to the case of desorption supported by diffusion (for example in a liquid such as PDMS or the superfluid 4He film). We have to note that a large number of particles solvated in the coating, which can diffuse in the bulk and latter easily desorbed from the surface makes a manifestation of this light induced diffusion-desorption effect ( LIDD effect \cite{TheEffect}) really spectacular. For example, a weak light from ambient lamp applied to a glass cell with a PDMS coating can produce a sodium vapor density that is some order of magnitude larger than the thermal value at room temperature \cite{Gozzini1993}.

This effect has received considerable attention in recent years due to its application as a light-controlled  atomic source. It has been successfully used to load magneto-optical atom traps \cite{Anderson2001,Atutov2003,Klempt2006,Zhang2009,Telles2010}, for production of a Bose-Einstein condensate (BEC) of  $^{23}$Na \cite{Mimoun2010}, generation of controlled Rb-vapor densities in photonic-band-gap fibers \cite{Londero2009,Bhagwat2009}. Its use has also been considered for atomic magnetometers, gyroscopes, and clocks \cite{Karaulanov2009,Bogi2009}.

There are several fields of research and application for which the ability to collect and then release by a pulsed light a large number of atoms or molecules can be a useful tool. For example, it is possible to collect by sorption on coated (or non coated) surface a large number of toxic or radioactive elements during tens of minutes and then release by photodesorption  all these elements to a detection volume in a short time of milliseconds by a powerful pulse of light. In a case of low loss of collected particles and high efficiency of photodesorption, a ratio of a pulsed signal, from burst of desorbed particles, to noise is proportional to the ratio of the collection time to the release time which can be of several orders of magnitude. This method could be used for enhancement of sensitivity of existed sensors for the trace detection of various elements (including toxic or radioactive ones) which is important to environmental applications, medicine or in geology. This might help, for example, to construct an efficient light-driven atomic source for a magneto-optical trap in a case of extremely low vapor density or very weak flux of atoms, such as artificial radioactive alkali atoms.

The purpose of this paper is to study the efficiency of the collection and realization by photodesorption effect of Rb atoms, a subject of interest for the applications discussed. Because the yield of LIDD effect is potentially larger than a photodesorption yield from a solid surface, we restrict ourselves to study the efficiency of this effect in the case of Rb atoms immersed in PDMS coating. We also demonstrate the importance of the curing or passivation of the coating on the cell walls with the Rb vapors in order to minimize the loss rate of collected Rb atoms because of the irreversible chemical bonding of the atoms by the film. We estimate the average number of bounces it takes to adsorb atoms back to the coating after desorbing light is off. The importance of long and narrow cell pumping tube, which minimizes a loss of collected atoms due to their leaking out of the cell is also demonstrated.
%
\section{Experimental setup}
\label{sec:2Setup}
\begin{figure}
  \includegraphics[width=0.47\textwidth, keepaspectratio]{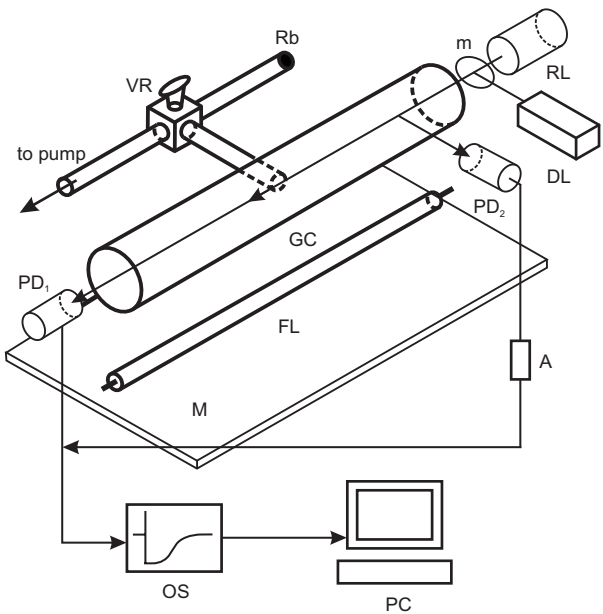}
  \caption{\label{fig1} Sketch of experimental setup. GC - glass cell, RL - resonance lamp, PD$_{1}$, PD$_{2}$ -photodetectors, DL - diode laser, VR - valve, Rb - piece of metallic rubidium, FL - flash lamp, M - aluminum mirror, m -moveable dielectric mirror, A - lock  in amplifier, OS - oscilloscope, PC - computer.}
\end{figure}

%
The main part of the setup is an adsorption cell made of a cylindrical Pyrex glass tube connected to a vacuum pump through a narrow glass exit or pump tube with a valve, an atomic vapor source of natural isotopic mixture of Rb (see Fig.~\ref{fig1}). The cell has internal diameter $R_{cell} = 0.75 \; \textnormal{cm}$ and length $L = 20 \; \textnormal{cm}$, the exit tube diameter is $r = 0.15 \; \textnormal{cm}$, length $l = 20 \; \textnormal{cm}$. The valve has three internal passes. They allow us either to connect the cell to the vacuum pump or to the Rb source or isolate the cell from both pump and source. The inner surface of one cell, exit tube and valve are covered by an organic coating. This coating is prepared from 3 \% solution of commercial PDMS liquid material (M.W. 170, 200, secondary standard, Aldrich Chemical Company. Inc.) in ether. The cell preparation is described, for example, in \cite{Atutov1999}.  Without proper control, the coating can be too thick: about 0.1 mm, and for high viscosity coating the atomic release time can be fairly long: up to $10^{4}$ seconds \cite{Atutov1999}. Note that the thickness of the coating was found to be roughly proportional to a concentration of PDMS. The film thickness of our cell, was determined via both microscopic and interferometric measurements. It was measured that the thickness of the coating is not uniform across the cell cylinder: on the cylinder top the thickness is 0.5 micron, on the bottom it is about 2 - 3 microns. This difference in thickness is attributed  to the liquid nature of the PDMS compound deposited on the cell walls.

Some techniques such as mass spectrometer, ionization followed by ions counting, laser fluorescent or adsorption spectroscopy can be used to study the evolution of the vapor density in the cell as a result of LIDD effect. In our experiment, the main measurements were performed using as a source of probing light  resonance discharge lamp filled by natural isotopic mixture of Rb atoms.

Several well known reasons for using a resonance lamp are the following. Besides being easy to operate, resonance lamps are cheap, using of them does not suffer from optical pumping through the hyperfine atomic levels  and, very importantly, their radiation can be extended over wide spectral regions, including those not covered by present laser radiations such as the UV region. Furthermore, the electron and ions kinetic energies of glow discharge are high enough to provide sufficient pressure of active atomic impurities in the lamp discharge zone, even when using materials with low volatility. 

However, for measurement of a small density of Rb atoms in the cell in vapor phase fluorescence detection was used. The fluorescence was exited by a free running diode laser which frequency tuned to Rb atom resonant transition of 780 nm. The laser radiation was sent by a moveable mirror. To avoid optical pumping through Rb hyperfine levels laser frequency was periodically swept across of four fine spectral lines of both $^{85}$Rb and $^{87}$Rb isotopes by a sinusoidal modulation of the diode laser current. The frequency of current modulation is 1.525 kHz. In this particular case, the lock-in amplifier was employed. The absorption or fluorescent signals were processed by a digital oscilloscope connected to the computer. The acquisition system allowed us to collect data with 0.1 ms resolution limit over several hundreds of minutes and to measure density variation of Rb vapor in a wide range. The absolute Rb vapor density in equilibrium with a metal vapor source was estimated from the temperature of the Rb metal drop, kept at room temperature \cite{Nesmeyanov1963}.
          
The LIDD effect of Rb atoms adsorbed in the PDMS film was studied using  flash lamp with  maximal fluence of the desorbing light from the lamps on the cell surface  $0.1 \; \textnormal{J/cm}^{2}$. The lamp was equipped by Al mirror placed near the lamp as shown on Fig.~\ref{fig1}. It reflected light through the lamp to the cell and hence provided a desorbing light fluence which was about 60 percent higher. The value of the fluence used was alternated by changing the distance between the cell and flash lamp. The cell was always completely illuminated by the flash light so as to exclude any influence of both atomic bulk and surface diffusion along the coating on the evolution of the vapor density in the cell. The exit tube, valve and Rb source were made non-transparent to exclude any disturbance of the Rb vapor inside of them by flash light.
%
\section{Passivation and evolution of atomic density in the vapor cell}
\label{sec:3Passivation}
%
We found that freshly-coated cell did not show any fluorescence from Rb atoms when the valve to the vapor source is opened, meaning that the life-time and the number of bounces of the atoms in the cell were very small \cite{Bouchiat1966}. This can be attributed to the fact that a fresh coating in vacuum  has a chemically active surface and bulk, probably because of a trace of gases such as oxygen or water adsorbed and mixed with the molecules of the coating. To minimize the residual chemical activity of the coating, we carried out a passivation (or curing) procedure \cite{Lu1997,Atutov2009} by use Rb vapor itself.

\begin{figure}
  \includegraphics[width=0.47\textwidth, keepaspectratio]{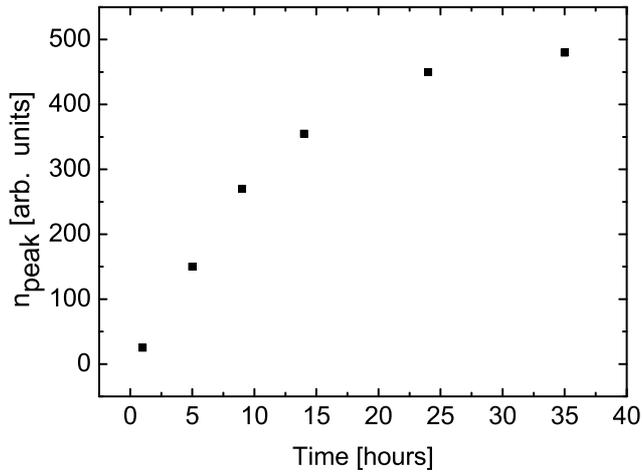}
  \caption{\label{fig2} Peak density (n$_{peak}$) of desorbed atoms versus time of passivation.}
\end{figure}

First of all, we pumped the cell continuously to obtain a residual-gas pressure of $10^{-8}$ mbar. It usually takes at least a few days to achieve these vacuum conditions in the cell. To start passivation, we heated the source of the rubidium atoms and opened the valve so that the pressure of the alkaline vapor in the cell was about $10^{-7}$ mbar. After this, we closed the valve and switched off the heating of the source so as to allow its temperature to reach the ambient value. Then we opened the valve and filled the cell with rubidium vapor during 30 s. Finally, we applied the flash light and recorded the transmitted light intensity of the probe lamp and evaluated the peak density of the desorbed atoms n$_{peak}$ inside the cell. Figure~\ref{fig2} shows how peak density npeak depends on passivation time. One can see that at the beginning of the passivation process, the peak density of desorbed atoms density is small. After about 40 hours of continuous passivation, the peak density approaches its limit. From our measurements, we deduced that, the increase in the peak density after and before passivation was approximately a factor $10^{4}$.

In order to check the ability of our cell to work and that the cell is completely passivated we performed a set of experiments to study the evolution of photodesorbed atoms and life-time of the atoms in the cell. The coating was exposed to Rb vapor during one week with the source kept at room temperature.

\begin{figure}
  \includegraphics[width=0.47\textwidth, keepaspectratio]{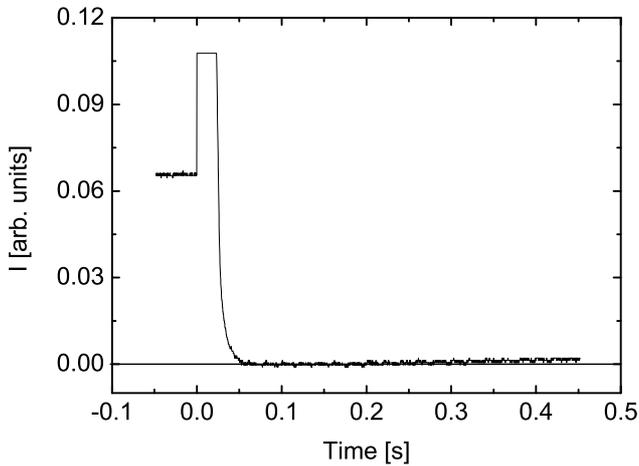}
  \caption{\label{fig3} Intensity of transmitted light (I) as a function of time at maximal fluence.}
\end{figure}

\begin{figure}
  \includegraphics[width=0.47\textwidth, keepaspectratio]{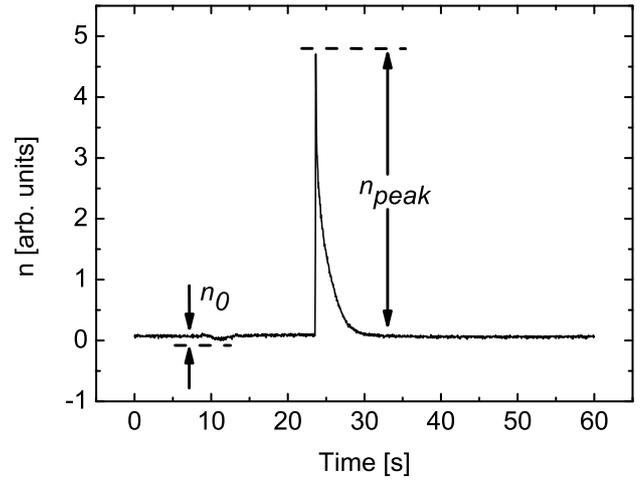}
  \caption{\label{fig4} Comparison of Rb vapor density (n$_{0}$) in the cell at room temperature to peak density of desorbed atoms (n$_{peak}$).}
\end{figure}

Figure~\ref{fig3} shows the intensity of the probe lamp light transmitted through the cell isolated from both pump and source versus time, with maximal $0.1 \; \textnormal{J/cm}^{2}$ fluence of the flash lamp. It is possible to see that at a high flash lamp fluence applied to completely passivated and saturated by Rb cell, the density of photodesorbed atoms because of the LIDD effect is so high that the cell minimum transmission comes close to zero, which means that after flash the Rb vapor becomes optically thick. We deduced the density of the photodesotbed atoms as a function of time from the transmitted signal by applying Beer's low and a typical result is demonstrated on Fig.~\ref{fig4}. Two arrows indicate the vapor density at equilibrium  n$_{0}$  by the fast closing and opening valve to the Rb source kept at room temperature. Figure~\ref{fig4} shows that the ratio of the collected and then photodesorbed atoms at maximum n$_{peak}$ is at least 50 times larger than the density at the equilibrium n$_{0}$. This ratio is smaller than those reported in \cite{Gozzini1993} because the equilibrium pressure of the Rb vapor at ambient temperature is higher than the Sodium pressure.

\begin{figure}
  \includegraphics[width=0.47\textwidth, keepaspectratio]{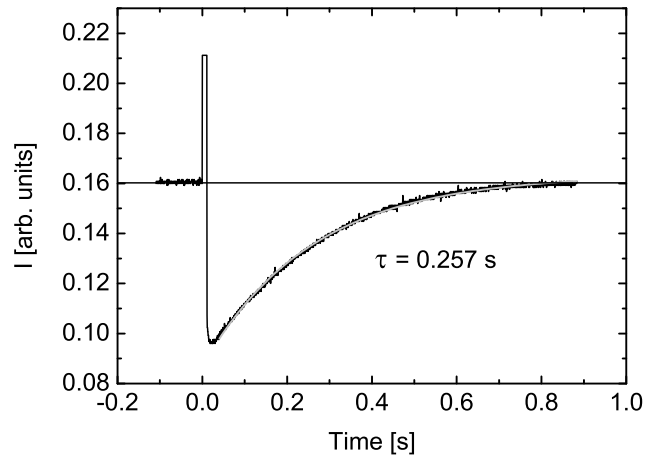}
  \caption{\label{fig5} Intensity of transmitted light (I) as a function of time at low fluence.}
\end{figure}

The life-time of the desorbed atoms in our cell has been studied using an intermediate flash lamp fluence of $0.03 \; \textnormal{J/cm}^{2}$. This relatively small fluence has been used in order to keep the vapor of desorbed atoms optically thin. Figure~\ref{fig5} shows the transmitted signal of the desorbed atoms as a function of time. This measurement  was done for the cell isolated from both  pump and  Rb source.  In an optically thin regime, the transparency of the cell is proportional to the atomic vapor density. The transmitted signal decreases very rapidly after the flash pulse. The spike at $t=0$ is due to the stray flash light hitting the photodetector. After the spike one can see that the density of the atoms in the cell increases in a time range of about 10 ms and then decay exponentially as a consequence of the readsorbtion of the atoms back to the coating. The best fit of the exponential decay shows a life-time to be equal to 0.257~s.

As it has been shown in \cite{Atutov2009} the life-time of atoms   in a coated cell can be written in the following form:
\begin{equation}
\label{Eq1}
\tau = \frac{\tau_{adsorb}\tau_{esc}}{\tau_{adsorb}+\tau_{esc}}
\end{equation}
where  $\tau_{adsorb}$ is a life-time of atoms inside the cell isolated from the pump before being lost due to adsorption back on the cell walls, and $\tau_{esc}$ represents the time which it takes to lose atoms through the exit tube to the reservoir with metallic Rb or to the pump.

In the case of cylindrical cell, assuming that the cell internal wall area much larger than the exit tube and two windows cross sections, we can express $\tau_{adsorb}$ as follows:
\begin{equation}
\label{Eq2}
\tau_{adsorb} = \chi\frac{R_{cell}}{2\overline{v}}
\end{equation}
and $\tau_{esc}$ can be written as:
\begin{equation}
\label{Eq3}
\tau_{esc} = \frac{3lR^{2}_{cell}}{2r^{3}\overline{v}}
\end{equation}
where parameter $\chi$ is interpreted as the average number of bounces it takes to adsorb atoms on the surface of the coating, $\overline{v}=\sqrt{\frac{8kT}{m}}$ is the average atomic thermal velocity at temperature $T$, and $m$ is the mass of the atom.

In the isolated cell the escape of the atoms is canceled, and the life-time of the atoms is dominated by physadsorption of the atoms on the cell walls only:
\begin{equation}
\label{Eq4}
\tau = \tau_{adsorb}
\end{equation}

Note that this calculations of life-time of atoms do not aspire to high accuracy and have a semiquantitative nature, but we suppose to estimate the order of magnitude of the number of bounces in cylindrical cell.

Taking into account Eq.~(\ref{Eq4}) and measured decay time of Rb atoms density, the number of bounces before readsorption on cell walls coated by PDMS is estimated to be about $10^{4}$. It is interesting to note that this number is consistent to the rise of the peak density of the photodesorbed atoms during time of passivation, which has been measured and discussed above, to the number of bounces of the Rb atoms in a Dry film coated cell \cite{Atutov2009} and to the number of bounces of the sodium atoms in a Paraffin coated cell \cite{Atutov1963}.

In comparison with the short adsorption time escape time is rather long. In fact,  for our cylindrical cell geometric parameters and Rb atom thermal velocity $v = 2.7 \cdot 10^{4} \; \textnormal{cm/s}$, by the use of Eq.~(\ref{Eq3}), we derived that it takes 12~s for Rb atom to leak out of the cell. We were not able to measure this escape time by photodesorption technique, which is more suitable for the short time measurements. Instead of the pulsed desorption technique we performed  an experiment in which the Rb vapor is pumped away from the cell.

\begin{figure}
  \includegraphics[width=0.47\textwidth, keepaspectratio]{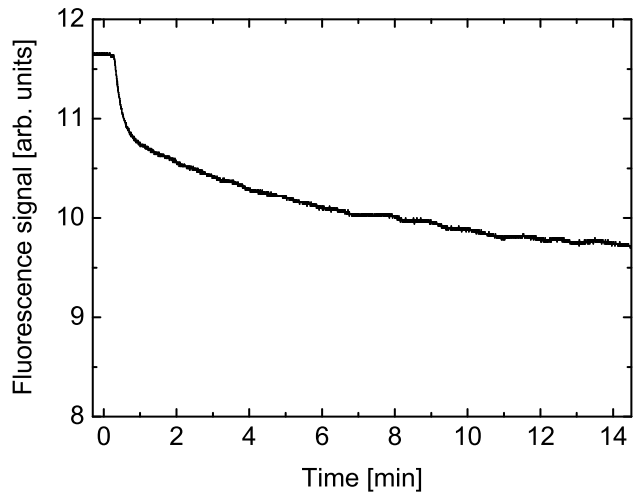}
  \caption{\label{fig6} Rb vapor density decay in the pumped cell.}
\end{figure}

We again exposed the coating to Rb vapor during one week. After the exposition a weak Rb fluorescence have appeared in the cell. To start measurement we blocked Rb source, at the same time opened the valve to ion pump  and monitored vapor fluorescence exited by the diode laser. Figure~\ref{fig6} illustrates how the fluorescence of the Rb vapor decays in the pumped cell. One can see that at the beginning of the pumping process Rb density drops during 11.5~s due to the leaking of the vapor out of the cell.
Because of the measured value of the escaping time is closed to the estimated one, we believe that there is no remarkable number of Rb atoms deposited on a liquid surface of PDMS film.

As can be seen, this fast exponential decay of the vapor density is followed by a long tail, which, as it was measured, lasted several hours. It is obvious, that this tail is due to a slow detachment of Rb atoms from the glass substrate together with a slow diffusion of Rb through the coating from the glass substrate towards to the coating surface. Because no desorbing light is applied, the surface atomic diffusion along the coating is excluded. Due to the incertitude of the atomic density value on the glass substrate surface and the atoms glass detachment rate, it is not possible to derive correct atomic diffusion coefficient in darkness in this experiment.
%
\section{Collection efficiency}
\label{sec:4Efficiency}
%
We provide a series of the efficiency measurements of the Rb LIDD effect. At the beginning of each measurement, we carefully cleaned the passivated coating from free Rb atoms, which can be photodesobed, This was done by the illumination of the cell by a 300~W lamp for the duration of one hour, keeping the cell continuously pumped and the Rb source closed. Then we isolated the cell from the pump and apply the flash lamp light in order to be sure that the coating has no free Rb atoms in the bulk.

To measure photodesorption efficiency we released a fixed number of atoms into the cell and determined the number of atoms which can be desorbed from the coating. It was done by a bit warmed Rb source. A higher temperature increases the density of the atoms in the source in vapor phase and thus increases the number of atoms, which can be released in the cell. We found also that the efficiency is roughly proportional to the total fluence in the used power range of the flash lamp. Thus, for this particular experiment we used three times more powerful flash lamp with maximal fluence of $0.3 \; \textnormal{J/cm}^{2}$.

\begin{figure}
  \includegraphics[width=0.47\textwidth, keepaspectratio]{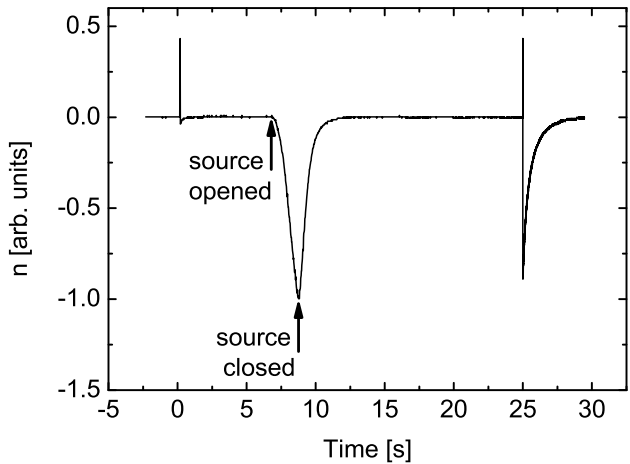}
  \caption{\label{fig7} Illustration of the efficiency measurement procedure, see in the text.}
\end{figure}

The procedure of the measurement is illustrated in Fig.~\ref{fig7}. Initially the cell is isolated from both pump and source. At 0 second the flash light is applied. Here there is no signal change, because the coating is free from Rb atoms. At 5.3~sec. the valve to the Rb source is opened and after 2~sec. closed. It is a gap in the line of the signal. Transmitted light signal drops because Rb atoms released into the cell, and then the signal increases to previous value due to that the atoms are adsorbed by the coating. After a storage time of 16 seconds the flash light was applied on the cell. Another gap appeared. Here, the transmitted light signal drops again but because atoms desorbed from the coating and it increases to the previous value because the photodesorbed Rb atoms are adsorbed again by the coating. Then after 30 minutes once more we applied flash light and detected transmitted signal. Then during 300 minutes, five another flashing were made with recording of transmitted signal.

\begin{figure}
  \includegraphics[width=0.47\textwidth, keepaspectratio]{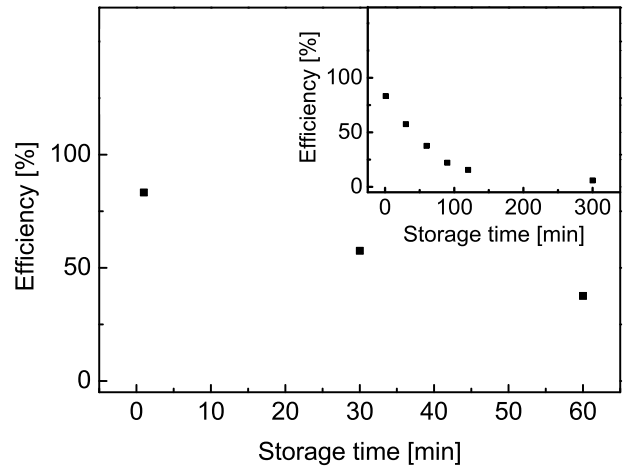}
  \caption{\label{fig8} Efficiency versus storage time, cell isolated from ionic pump. The inset shows how efficiency decays during 5 hours storage time.}
\end{figure}

Due to the linear response of the photodetector and in the optically thin vapor case,  the integral of any gaps is proportional to the adsorption of the probe light in the cell i.e. to the total number of  atoms in a vapor phase in the cell. To determine the collection efficiency, we compared part of the integral of the first gap for time when the valve was opened, to the integral of the second gap. The ratio of these two integrals gives a value that is proportional to the efficiency of the adsorption/photodesorption process. It was determined that 85 percent collected chemically active Rb atoms stored during 16 seconds in the closed cell can be desorbed by single flash of the photographic lamp. The efficiency as a function of storage time for the closed cell is shown in Fig.~\ref{fig8}. One can see that the efficiency slowly decreases in time because of slow loss of the atoms imbedded in the coating. The characteristic time of the decay is equal to 60 minutes.

The decay of the efficiency in the case of isolated cell is due to a loss in the cell of the collected atoms. This loss corresponds to the fact that the vapor pressure measured in the passivated cell in dark is always slightly lower than the vapor pressure inside of the cell connected to the Rb source (see also \cite{Seltzer2010}). We believe that this is because of the existence of the continuous loss of atoms via reversible physical adsorption of the atoms onto the glass substrate of the coating, as was discussed above. Atoms from the source (where the atomic density is maximal) diffuse through the exit tube to the cell volume and then, after some bounces between the cell walls, they are finally physadsorbed by the coating and thus they can diffuse within the bulk volume of the film. These atoms then diffuse  deep inside the coating (where the density of atoms is minimal) and they became trapped by glass substrate of the organic film. Thus, the continuous gradient of the density of the atoms from the source to the substrate causes continuous flux of atoms and it takes a long time for the cell to achieve a truly steady state. For instance, we have equilibrium density $\textnormal{n}_{0}$  in the source $5.7 \cdot 10^{9} \; \textnormal{cm}^{-3}$, (at the temperature $20^{\circ}\textnormal{C}$, \cite{Nesmeyanov1963}), density gradient from the source to the cell $n/l = 2.85 \cdot 10^{8} \textnormal{cm}^{-4}$, for exit tube radius 0.15~cm a flux of atoms is $\sim 10^{11} \; \textnormal{s}^{-1}$ . For this flux and for the internal surface of the cell equals to $90 \; \textnormal{cm}^{2}$, one can evaluate that it takes about $10^{7}$ seconds or about one year to produce one atomic monolayer on the coating substrate.

Rubidium atoms could react with a coating, for example, they could be also trapped by oxygen which is not shielded by methyl groups in silicon-oxygen backbone of PDMS molecules, leaving fragments of the molecules in the cell. The existence of the reaction between the coating and atoms was studied and discussed in details in \cite{Stephens1994}. The possibility of the reactions between the coating and atoms was also demonstrated \cite{Yi2008} for the case of Rb in a cell coated by octadecyltrichlorosilane OTS  compound. In this work, to monitor the occurrence of any chemical reactions between the rubidium and the walls, the composition of the rest gas in the cell was analyzed with a mass spectrometer. Recorded mass spectrum of the background gas inside of the OTS cell indicates some weak reactions between the coating and the rubidium atoms. Similar fragments were found in cells with dichlorodimethylsilane coating after exposure to Rb \cite{Camparo1987}.

\begin{figure}
  \includegraphics[width=0.47\textwidth, keepaspectratio]{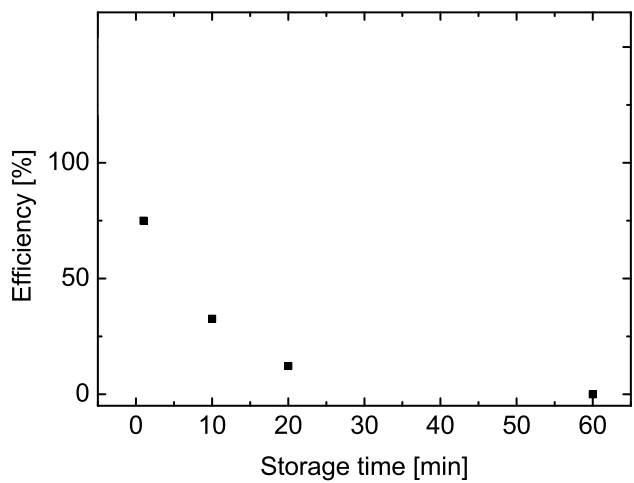}
  \caption{\label{fig9} Efficiency versus storage time, cell connected to ionic pump.}
\end{figure}

We measured the desorption efficiency for the pumped cell. This particular case is interesting for the possible application of LIDD effect for collection and trapping of artificial radioactive alkali atoms in magneto optical trap, because in this case a MOT cell is permanently connected to accelerator as a source of trappable atoms. The procedure of measurement was the same, only that the valve to the pump remained opened during all measurements. Figure~\ref{fig9} shows the efficiency decay in time. One can see that the number of the released atoms and efficiency decay faster than in the case of isolated cell. It can be determined that for the pumped cell the characteristic decay time is 12.4 minutes. It was also determined that 75 percent collected chemically active Rb atoms stored during 16 seconds in the pumped cell can be desorbed by single flash of the lamp. A bit smaller efficiency in the case of pumped cell, with respect to the case of isolated cell, is the result of pumping of atoms out of coating during 16 second storage time.

Form and characteristic decay time of the curve shown on Fig.~\ref{fig8}, is different from one on Fig.~\ref{fig6} because of the decay of Rb atoms immersed in the bulk of the coating, shown on Fig.~\ref{fig9}, is in principle different process than vapor fluorescence decay. Decay time of Rb solved in the coating (Fig.~\ref{fig9}) is much smaller in comparison to the time on Fig.~\ref{fig6}. This can be explained as different exposition time of the coating to the Rb vapor: one week in the vapor pumping out the cell experiment and two seconds in the last experiment. One week exposition time seems to be enough to make a non-negligible amount of Rb deposited on the glass substrate surface, two seconds might be not. Anyway, in the case of negligible number of the Rb atoms deposited on the glass substrate, taking into account decay time of 12.4 minutes shown on Fig.~\ref{fig9}, for coating thickness of 3 microns, one can estimate bulk diffusion coefficient in the dark for our PDMS film, which is about $10^{-10} \; \textnormal{cm}^{2}\textnormal{/s}$. It is about five orders of magnitude greater than $1.2 \pm 0.7 \cdot 10^{-5} \; \textnormal{cm}^{2}\textnormal{/s}$ measured in \cite{Kasprowicz2004}, unfortunately for unknown coating parameters such as coating thickness and PDMS viscosity, etc.

It is noted above that the life-time of photodesorbed atoms in the cell is a combination of readsorption and escape time. In our case of the cell with long and narrow pumping tube the escape time is greater than the readsorption time. On the other hand all desorbed atoms have enough time to be readsorbed by the coating rather than leak out of the cell. This means, that the flash light, in such experiment, appears as a tool for monitoring percentage of collected atoms remains in bulk of the coating kept in a dark during the storage time.

On the basis of the efficiency measurement, one is led to give recommendations for the efficient LIDD processes, for example in a MOT cell fed by a radioactive ionic beam, as it follows. After the radioactive ion beam is switched on, the ions come inside the cell and impinge on the neutralizer, stick to its surface for a short time, become neutralized and finally they are desorbed from the neutralizer and released into the cell volume. Atoms can also be injected directly into the cell in neutral form. In both cases, atoms start to fill the cell. After some number of bounces atoms become adsorbed by the coating. As it was shown, a complete adsorption of the atoms in our cell takes about hundreds of milliseconds. If the adsorption time is much smaller than the atomic escape time, the most of the atoms are absorbed by the coating rather than these atoms get lost by their leaking back trough the exit tube. Thus, a small adsorption time in a comparison with a leaking time together with a low loss of adsorbed atoms inside of the coating have a crucial importance for a high efficiency of the collected/desorbed atomic process. This requirement can be expressed as:
\begin{equation}
\label{Eq5}
\tau_{leak} \gg \tau_{adsorb}
\end{equation}
by using Eqs.~(\ref{Eq2},~\ref{Eq3}) we obtain:
\begin{equation}
\label{Eq6}
\frac{3lR^{2}_{cell}L}{2r^{3}\overline{v}} \gg \chi\frac{R_{cell}}{2\overline{v}}
\end{equation}
or
\begin{equation}
\label{Eq7}
\frac{3lR_{cell}L}{\chi r^{3}\overline{v}} \gg 1
\end{equation}
This estimation led one to the following conclusion. To have a high efficiency of the collection/desorption process, cell should have as large as possible internal volume. To minimize the loss rate from the cell, the exit tube through which the atoms enter in the cell should be small enough in diameter and long enough in length to minimize the leak rate. To minimize the loss of atoms via chemical adsorption by the inner surface of the cell and tube, the inner surface and exit tube have to be covered by any non stick coating (such as PDMS ), which have to be properly passivated. The temperature of the cell should be as low as possible in order to decrease number of the bounces of the atoms in the cell. Note, that the temperature anyway should be greater than photodesorption activation temperature in order to keep the atoms photodesorption rate high enough \cite{Atutov1999}. The last requirement is different to those discussed in \cite{Atutov2009}. This is due to that we consider in this experiment atoms solved in the coating but not atoms in vapor phase in the MOT cell.
%
\section{Conclusion}
\label{sec:5Conclusion}
%
In conclusion, we have demonstrated that efficient collection/photodesorption of chemically active Rb atoms is possible in a coated cell. It was found that 85 percent collected Rb atoms and stored during 16 seconds in the closed cell, 75 percent in the pumped cell can be desorbed by single flash of the lamp. The number of stored atoms decays with a characteristic time of 60 min in isolated cell and with a time 12.4 minutes in a pumped cell. We believe that this efficient method of collection and fast realization of atoms or molecules could be used for enhancement of sensitivity of existed sensors for the trace detection of various elements (including toxic or radioactive ones) which is important to environmental applications, medicine or in geology. The effect might help to construct an efficient light-driven atomic source for a magneto-optical traps in a case of extremely low vapor density or very weak flux of atoms, such as artificial radioactive alkali atoms. Our cell glass walls is not transparent to UV part of the lamp radiation, nevertheless a reasonable high efficiency was obtained in the experiment. Since it has been demonstrated that the UV wavelength of the desorbing light plays a non-negligible role (see for instance Ref. \cite{Atutov1999,Gozzini1993,Klempt2006,Telles2010,Karaulanov2009}), change of cell materials from a glass to a quartz would also further improve the desorption efficiency. In a near future, we plane to build a quartz vapor cell for detection of a trace of Mercury in an environment. This can be made by the use of a Hg$^{0}$ discharge probe lamp, which produces a radiation resonant to Hg$^{0}$ atoms optical resonant transition ($6s ^{1}S_{0} \rightarrow 6p ^{3}P_{1}$ transition at 253.7 nm) and a powerful flash lamp.
%
\section{Acknowledgments}
%
We would like to thank Prof. A.~Shalagin, A.~Kuch'yanov  and T.~Lukina for their helpful discussions. This work is supported by the integration project of the SB RAS \textnumero 17.
%


\begin{thebibliography}{38}
\bibitem{TheEffect}
The effect is often impropriety called light - induced atomic desorption - LIAD effect. As it was experimentally demonstrated that not only atoms but also molecules and clusters can effectively desorbed by this diffusion supported photodesorption process.
\end{thebibliography}
\end{document}